\documentclass[showpacs,preprintnumbers,amsmath,amssymb]{revtex4}
\usepackage{graphicx}
\usepackage{dcolumn}
\usepackage{bm}
\usepackage{braket}

\begin{document}
\def\a{ a}
\def\ak{ a^\dagger}
\def\bra#1{\langle #1|}
\def\ket#1{| #1\rangle}
\def\skal#1#2{\langle #1| #2\rangle}

\def\bran#1{\widetilde{\langle{ #1|}}}
\def\ketn#1{\widetilde{|#1\rangle}}

\def\norma#1{e^{-{{{\vert #1 \vert} \over 2}}^2}}
\def\normaa#1{e^{{{{\vert #1 \vert} \over 2}}^2}}
\def\tay#1#2#3{{#1^{\!^{(\!#2\!)}}\!\!(#3)}}
\def\tayf#1#2#3{{{{#1^{\!^{(\!#2\!)}}\!\!(#3)}}\over{#2!}}}
\def\pocjed#1{\begin{equation}\label{eq:#1}}
\def\jed#1{(\ref{eq:#1})}

\title{Analytic functions of the annihilation operator}

\author{Aleksandar Petrovi\'c}
\email{a.petrovic.phys@gmail.com}
\address{M.Gorkog 27, 11000 Belgrade, Serbia}


\begin{abstract}
A method for construction of an non-entire function $f$ of the annihilation operator $\a$ is given for the first
time. $f(z)$ is analytic on some compact domain that does not separate the complex plane. A new form of the
identity is given, which is well suited for the function's domain. Using Runge's polynomial approximation theorem,
such a function $f$ of the annihilation operator is defined on the whole domain. The constructed operators are
given in terms of dyads formed of Fock states.
\end{abstract}


\maketitle

\section{Introduction}
The annihilation operator in Hilbert space is operator $\a$ which
satisfies relation \pocjed{komutator}
 [ \a,\ak ]=1.
\end{equation}
A self-adjoint Number operator $N$ and the corresponding Fock
states are constructed using $\a$:

\pocjed{operator_br_cestica} N=\ak\a, \ \ \ \ \  N\ket n =n\ket n
,\ \ \ n=0,1,2,3,...
\end{equation}

\noindent
Eigenstates of $\a$ are called coherent states:

\pocjed{koherent1} \a\ket{\alpha} =\alpha \ket{ \alpha }, \ \ \ \
\ \ \ket{\alpha}=\norma \alpha \sum\limits_{n=0}^\infty
{{\alpha^n}\over{\sqrt{n!}}} \ket n ,\ \ \ \alpha\in \mathbb{C}.
\end{equation}

\noindent
The coherent states are denoted with Greek letters
($\ket \alpha$,$\ket \beta$,$\ket \gamma$), and the Fock states
with Latin letters ($\ket l$,$\ket k$,$\ket n$,$\ket m$).

\noindent
The identity can be formed using eigenvectors $\ket
\alpha$: \pocjed{stara_jedinica} I={1\over\pi}   \int  d^2 \alpha
\ket \alpha \bra \alpha .
\end{equation}
$\a$ is not a normal operator, so The Spectral Theorem cannot be
applied to construct $f(\a)$. Nevertheless, every entire function
of the operator $\a$ can be represented in a form analogous to the
resolution of a normal operator with respect to its projective
measure: \pocjed{razlaganje3} f(\a)={1\over\pi} \int d^2\alpha
f(\alpha)\ket \alpha \bra \alpha .
\end{equation}
Contrary to the normal operator case, non-entire functions of $\a$ cannot be represented in such a way
\cite{arsa},\cite{arsa1}. Previous unsuccessful attempts to use $\ln\a$ were pointed out in \cite{arsa}. Related
questions concerning this issue were also studied by Perelomov\cite{peleromov}. The underlying problem is the fact
that until now non-entire functions of the annihilation operator have not been studied in detail, and some
straightforward assumptions arising from resolution \jed{razlaganje3} have lead to erroneous conclusions. In this
paper, we construct a function $f(\a)$ of the annihilation operator, where $f$ is analytic on a compact domain
that does not separate the complex plane. The operators constructed here are given in terms of dyads formed of
Fock states.

\section{Results}
\subsection{New identity resolution}

Standard resolution of identity \jed {stara_jedinica} is not
suitable for non-entire functions, so a new resolution of identity
is given. To do that, non-normalized coherent states (NCS) are
defined:

\pocjed{nenormirana_stanja}
 \ketn \alpha=e^{\alpha\ak} \ket 0 =  \normaa \alpha \ket \alpha .
\end{equation}
Translation operator for NCS  is: \pocjed{translacioni_operator}
T(\beta)=e^{\beta\ak} ,\;\;T(\beta)\ketn \alpha =\ketn
{\alpha+\beta}  ,\ \ \  \alpha , \beta \in \mathbb{C}.
\end{equation}
Now, an identity whose eigenstates are on the circle of radius R
centered at the origin can be formed from NCS :

\pocjed{nova_jedinica1} I=-i\oint\displaylimits_{\vert \gamma\vert=R} {{d\gamma}\over{\gamma}} \ketn \gamma \bran
\gamma \; J,
\end{equation}

\noindent
where

\pocjed{nova_jedinica3} J={1\over{2\pi}} \sum\limits_{n=0}^\infty {n!} \ket n \bra n  .
\end{equation}

\subsection{Runge's approximation theorem and functions of the annihilation operator}

In this subsection, $f(\a)$ where $f$ is an analytic function defined on a compact domain $\Omega$ that does not
separate the complex plane, is constructed. Two classes of such functions are considered. The first class consist
of functions $f$ defined on domains containing $0$: $0\in\Omega$. The second class consists of functions which are
not defined at $0$: $0\not\in\Omega$. A convenient theoretical tool for this is Runge's polynomial approximation
theorem. \vskip0.5cm

\noindent {\bf Theorem:}  If $f$ is an analytic function on a compact domain $\Omega$ that does not separate the
complex plane, then there exists a sequence $P_l(z)$ of polynomials such that  converges uniformly to $f(z)$ on
$\Omega$ \cite{saff},\cite{markushevic},\cite{ahlfors}:
 \noindent

\pocjed{runge1} f(z)=\sum\limits_{l=0}^\infty P_l(z-z_0)=\sum\limits_{l=0}^\infty\sum\limits_{k=0}^{d_l} c_k^{(l)}
(z-z_0)^k , c_k^{(l)}\in C, z,z_0 \in \Omega.
\end{equation}

\noindent Let us consider the first class of analytic function
($0\in\Omega$) and let

\pocjed{runge2}
f(z)=\sum\limits_{l=0}^\infty\sum\limits_{k=0}^{d_l} c_k^{(l)}
z^k , z \in \Omega
\end{equation}
be an expansion of $f$ in a series of polynomials. Then
\jed{runge2} and \jed {nova_jedinica1} can be used to construct
the function $f(\a)$ of annihilation operator:

\pocjed{runge3} f(\a)=\sum\limits_{l=0}^\infty\sum\limits_{k=0}^{d_l} c_k^{(l)} \a^k \cdot I= -i
\sum\limits_{l=0}^\infty\sum\limits_{k=0}^{d_l} c_k^{(l)} \oint\displaylimits_{\vert \gamma\vert=R}\!\!\! d\gamma
\gamma^{k-1} \ketn{\gamma} \bran\gamma \; J.
\end{equation}

\noindent
Using definition \jed {nenormirana_stanja}  of
$\ketn\gamma$, and after performing integration, $f(\a)$ becomes:

\pocjed{Mittag31} f(\a)=\sum\limits_{l=0}^\infty
\sum\limits_{n=0}^\infty \sum\limits_{m=n}^{d_l+n} c_{m-n}^{(l)}
\sqrt{{m!}\over{n!}} \;\;    \;\;\ket n \bra m\; ,
\end{equation}
\noindent
which for $\alpha\in\Omega$ gives:

\pocjed{Mittag4} f(\a)\ket\alpha=f(\alpha)\ket\alpha .
\end{equation}

\noindent Let us now examine the second class of $f$. Again we can
use \jed{runge1}, but requiring $z_0\not=0,z_0\in\Omega$. However,
a repetition of the above procedure using \jed{nova_jedinica1}
leads to a divergence problem. To resolve this issue, a different
resolution of identity, one involving translation
\jed{translacioni_operator} is chosen:

\pocjed{nova_jedinica5} I=e^{z_0 \ak} I e^{-z_0 \ak} .
\end{equation}

\noindent If $I$ in the right hand side of equation
\jed{nova_jedinica5} is substituted with \jed{nova_jedinica1}, and
using \jed{translacioni_operator} we get:

\pocjed{nova_jedinica6} I=-i\oint\displaylimits_{\vert \gamma\vert=R} {{d\gamma}\over{\gamma}} \ketn {\gamma+z_0}
\bran \gamma \; J \;e^{-z_0 \ak} \ .
\end{equation}

\noindent Now we  apply the operator, obtained by polynomial
approximation, to the identity \jed{nova_jedinica6}:

\pocjed{Mittag2} f(\a)=\sum\limits_{l=0}^\infty\sum\limits_{k=0}^{d_l} c_k^{(l)} (\a-z_0)^k \cdot I=
-i\sum\limits_{l=0}^\infty\sum\limits_{k=0}^{d_l} c_k^{(l)} \oint\displaylimits_{\vert \gamma\vert=R}\!\!\!
d\gamma \gamma^{k-1} \ketn{\gamma+z_0} \bran\gamma \; J\;e^{-z_0 \ak} .
\end{equation}

\noindent Again, using definition \jed {nenormirana_stanja}  of
$\ketn\gamma$, and after performing integration, $f(\a)$ becomes:

\pocjed{Mittag3}
f(\a)=\sum\limits_{l=0}^\infty\sum\limits_{k=0}^{d_l} c_k^{(l)}
\sum\limits_{n=0}^\infty \sum\limits_{m=k}^{n+k} {n\choose m-k}
\sqrt{{m!}\over{n!}} \;\; z_0^{n-m+k}   \;\;\ket n \bra m\;e^{-z_0
\ak} ,
\end{equation}

\noindent
which for $\alpha\in\Omega$ gives:

\pocjed{Mittag3a} f(\a)\ket\alpha=f(\alpha)\ket\alpha .
\end{equation}

\noindent We can simplify \jed {Mittag3}, collecting coefficients
at dyads, as

\pocjed{Mittag5} f(\a)= \chi e^{-z_0 \ak},
\end{equation}

\begin{displaymath}
\chi = \sum\limits_{l=0}^\infty  \sum\limits_{n=0}^\infty
\sum\limits_{m=0}^\infty \chi_{nm}^{(l)} \ket n \bra m ,
\end{displaymath}

\begin{displaymath}
\chi_{nm}^{(l)} = \sqrt{{m!}\over{n!}} \sum\limits_{k=p}^{s_l}
c_k^{(l)}   {n\choose m-k} \; z_0^{n-m+k} \;\;\;
,p=\max\{0,m-n\},\;\;s_l=\min\{m,d_l\} .
\end{displaymath}

\section{Discussion and conclusion}

In this paper, for the first time an expression for a non-entire function $f(\a)$ of the annihilation operator is
obtained. $f(z)$ is an analytic function on an compact domain $\Omega\subset \mathbb{C}$ that does not separate
the complex plane.

\noindent A very significant application of our result is its use in a construction of $\ln \a$. Since $\ln z$ and
$1/z$ can be defined on domain which satisfies previous conditions, our method allows us to obtain $\ln \a$ and
$1/\a$. Since, formally, $[\ln \a ,\a^\dagger]=1/\a$, it follows

\pocjed{faza} [\a^\dagger a,-i\ln \a]=i.
\end{equation}
Operator $-i\ln \a$ is conjugate to Number operator. It
 is not self-adjoint, but due to commutation relation \jed{faza} it can
serve as a good starting point for construction of the Phase
Operator \cite{arsa},\cite{arsa1}. Considering the significance of
this result, it will be topic of a separate paper.

\acknowledgments The autor would like to thank Drs M.Arsenovi\'c,
D.Arsenovi\'c, D.Davidovi\'c and J.Ajti\'c for their suggestions
and help in preparation of this manuscript.

\bibliography{apssamp}

\end{document}